\documentclass[11pt,twocolumn]{aastex63}
\usepackage{url}
\usepackage{gensymb}
\usepackage{textcomp}
\usepackage{xspace}
\usepackage{natbib,amsmath} \usepackage{graphicx}
\usepackage{hyperref} \usepackage{float} \usepackage{subfigure}
\usepackage{tabularx}
\providecommand{\e}[1]{\ensuremath{\times 10^{#1}}}
\usepackage{bm}
\newcommand{\um}{\textmu m\xspace}

\newcommand{\ang}{\AA\xspace}

\newcommand{\g}{\mathrm{g}}
\newcommand{\K}{\mathrm{K}}

\newcommand{\s}{\mathrm{s}}

\newcommand{\p}{\mathrm{p}}

\usepackage{array}

\begin{document}
\title{More Evidence for Variable Helium Absorption from HD 189733b}

\correspondingauthor{Michael Zhang}
\email{mzzhang2014@gmail.com}

\author[0000-0002-0659-1783]{Michael Zhang}
\affiliation{Department of Astronomy, California Institute of Technology, Pasadena, CA 91125, USA}

\author[0000-0001-9207-0564]{P. Wilson Cauley}
\affiliation{Laboratory of Atmospheric and Space Physics, University of Colorado Boulder, Boulder, CO 80309}

\author[0000-0002-5375-4725]{Heather A. Knutson}
\affiliation{Division of Geological and Planetary Sciences, California Institute of Technology}

\author[0000-0002-1002-3674]{Kevin France}
\affiliation{Laboratory of Atmospheric and Space Physics, University of Colorado Boulder, Boulder, CO 80309}

\author[0000-0003-0514-1147]{Laura Kreidberg}
\affiliation{Max Planck Institute for Astronomy, Heidelberg, Germany}

\author[0000-0002-9584-6476]{Antonija Oklop{\v c}i{\'c}}
\affiliation{Anton Pannekoek Institute of Astronomy, University of Amsterdam, Science Park 904, 1098 XH Amsterdam, Netherlands}

\author[0000-0003-3786-3486]{Seth Redfield}
\affiliation{Astronomy Department and Van Vleck Observatory, Wesleyan University, Middletown, CT 06459, USA}

\author[0000-0002-7260-5821]{Evgenya L. Shkolnik}
\affiliation{School of Earth and Space Exploration, Arizona State University, Tempe, AZ 85287, USA}

\begin{abstract}
We present a new Keck/NIRSPEC observation of metastable helium absorption from the upper atmosphere of HD 189733b, a hot Jupiter orbiting a nearby moderately active star.  We measure an average helium transit depth of $0.420 \pm 0.013$\% integrated over the [-20, 20] km/s velocity range.  Comparing this measurement to eight previously published transit observations with different instruments, we find that our depth is 32\% (9$\sigma$) lower than the average of the three CARMENES transits, but only 16\% (4.4$\sigma$) lower than the average of the five GIANO transits.  We perform 1D hydrodynamical simulations of the outflow, and find that XUV variability on the order of 33\%--common for this star--can change the helium absorption depth by up to 60\%, although a more typical change is 15\%.  We conclude that changes in stellar XUV flux can explain the observational variability in helium absorption, but that variability in the stellar He line cannot be excluded.  3D models are necessary to explore other sources of variability, such as shear instability and changing stellar wind conditions.
\end{abstract}


\section{Introduction}
\label{sec:introduction}
HD 189733b is currently one of the most extensively studied exoplanets in the literature.  Its exceptional observational favorability has made it a popular target for atmospheric characterization studies.  It is a large (1.1 $R_J$) planet orbiting a somewhat small K dwarf (0.78 $R_\Sun$) on a relatively tight orbit (P=2.2 d), and is located just 20 pc from Earth \citep{addison_2019,rosenthal_2021}.  Since its discovery in 2005 \citep{bouchy_2005}, it has been observed in wavelengths ranging from the X-ray to the mid-infrared, using a wide variety of observational approaches.  HD 189733b is currently the only exoplanet with a possible X-ray transit detection \citep{poppenhaeger_2013}, and was the first planet with a measured infrared phase curve \citep{knutson_2007}.  It has an extensively characterized transmission and emission spectrum (e.g. \citealt{sing_2011,mccullough_2014,morello_2014}), which provide us with some of the most precise constraints on its atmospheric composition (e.g. \citealt{zhang_2020b,harrington_2022}).  This planet has been the focus of a diverse array of theoretical studies seeking to explain its observed properties, including its atmospheric circulation patterns (e.g. \citealt{showman_2009}) and the nature of its atmospheric aerosols (e.g. \citealt{steinrueck_2021}).

HD 189733b is also well suited for mass loss studies, for many of the same reasons that make it favorable for observations of the lower atmosphere.  Atmospheric mass loss is an important phenomenon that could reshape exoplanet demographics, potentially turning mini-Neptunes into smaller and denser super-Earths, as well as creating the Fulton gap \citep{fulton_2017,fulton_2018} and the Neptunian desert (e.g. \citealt{szab_2011,fulton_2018}).  However, there are many theoretical uncertainties in mass loss models that make exoplanet atmospheric evolution difficult to model, including the effects of magnetic fields \citep{adams_2011} and of interaction with the stellar wind.  The 10833 \ang line has long been proposed as a possible probe of exoplanet atmospheres \citep{seager_2000,oklopcic_2018}, but observations have only recently succeeded in detecting exoplanetary absorption in this line \citep{spake_2018,nortmann_2018,allart_2019}.  It has been suggested that K-type stars have the most suitable stellar spectrum for populating the metastable helium state  \citep{oklopcic_2019,poppenhaeger_2022}.  HD 189733b, aside from being large, nearby, and highly irradiated, orbits a K-type star.

In 2016--2017, \cite{salz_2018} observed three transits of HD 189733b with CARMENES and detected clear He I absorption in all three, with consistent depths.  \cite{guilluy_2020} observed five transits with GIANO and obtained inconsistent depths indicating that HD 189733b's outflow may be varying in time.  In this paper, we present a new transit observation with Keck/NIRSPEC.  Section \ref{sec:data} describes the data acquisition, Section \ref{sec:analysis} the data analysis, Section \ref{sec:comparison} compares our results to prior work, Section \ref{sec:model} uses a 1D hydrodynamic model to explore how much the varying high-energy stellar flux should change the signal, and Section \ref{sec:conclusion} summarizes our conclusions.

\section{Observations}
\label{sec:data}
We observed a transit of HD 189733b with Keck/NIRSPEC on July 14, 2020, from 08:28 UTC to 14:41 UTC, using the NIRSPEC-1 filter and the 0.288x12$"$ slit.  These observations cover a wavelength range of 0.947--1.121 \um at a resolution of R=37,500.  Over the course of these 6 hours, we took 252 exposures in a ABBA nod pattern, with each exposure taking slightly over a minute and each ABBA pattern taking roughly 5 minutes.  These data cover 1.0 h of pre-transit baseline, the 1.8 h transit, and 3.4 h of post-transit baseline.  The airmass decreased from 1.2 at the beginning of the night to 1.0 around 0.5 h after mid-transit, before rising again to 1.66 at the end of the observations. The resulting SNR per spectral pixel in the continuum surrounding the He I line ranged from 150--280, with a SNR peak 2 hours after transit, and an unfortunate SNR trough coinciding with mid-transit.

Two of the raw spectra have multiple traces due to telescope nodding errors: one during egress, and one an hour after the end of egress.  We exclude these spectra and their nod companions from the analysis, for a total of four excluded reduced spectra.

The NIRSPEC observation was intended to be simultaneous with a Ly$\alpha$ observation by the Hubble Space Telescope's COS spectrograph (GO 15710).  Unfortunately, HST failed to acquire the guide star, so we did not collect any Ly$\alpha$ data.

\section{Analysis}
\label{sec:analysis}
To analyze these observations, we use the methods outlined in \cite{zhang_2022b}.  We make a master dark by stacking 5 darks with an exposure time of 1.5 s and 5 coadds.  We make a master flat by subtracting the master dark from each flat and stacking the 29 flats.  We create four difference images out of each $A_1B_1B_2A_2$ nod group: $A_1 - B_1$, $B_1 - A_1$, $B_2 - A_2$, $A_2 - B_2$.  This subtracts off bias, dark current and skyglow.  We then use a custom variant of optimal extraction to extract the 1D spectrum, while masking both bad pixels identified in the dark and flat processing stages, and cosmic rays identified by the optimal extraction algorithm itself.  Afterwards, we generate a template spectrum by combining a PHOENIX model \citep{husser_2013} with a telluric model, accounting for the radial velocity of the star relative to Earth.  We use this template spectrum to obtain a wavelength solution for order 70 (containing the helium line) in each individual exposure and then run \texttt{molecfit} to correct for tellurics.  This last step is not strictly necessary for these data because--due to a fortuitous combination of the star's radial velocity and Earth's orbital velocity--there are no telluric lines (with the possible exception of microtellurics) that overlap with the helium line.

Having obtained wavelength calibrated and telluric corrected spectra, we interpolate all spectra onto a common wavelength grid with a uniform logarithmic spacing of $\lambda/110,000$ and a range of 10,810--10,850 \ang.  The resolution of 110,000 is chosen so that the spacing of the grid is equal to that of the spectral pixels at 10833 \ang.  We remove fringing by applying a notch filter twice, using the same parameters as in \cite{zhang_2021}.  We divide each spectrum by the continuum, take the logarithm of the entire spectral grid, and subtract the mean of every row and column from that row and column.  The end result is a $N_{obs} \times N_{wav}$ grid of numbers representing the relative deviation of a pixel from the mean for that row and column.

\begin{figure*}[ht]
\centering
  \includegraphics
    [width=0.8\textwidth]{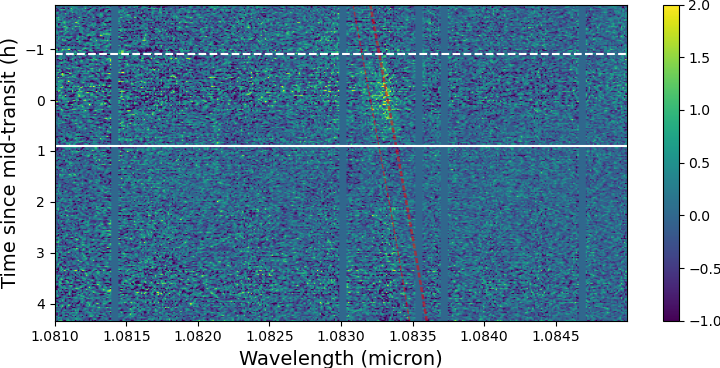}
    \caption{Excess absorption in percent, as a function of time and wavelength.  The diagonal red lines represent the wavelengths of the helium lines.  The horizontal white light represent the beginning of ingress and end of egress in white light.}
\label{fig:excess_2D}
\end{figure*}

\begin{figure}[ht]
  \includegraphics
    [width=0.5\textwidth]{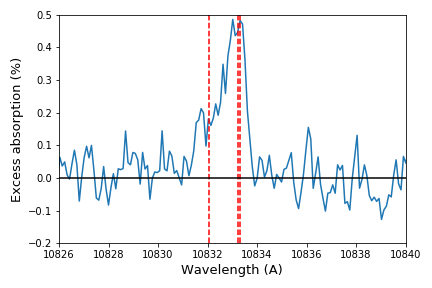}
    \caption{Average in-transit excess absorption spectrum from beginning of ingress to end of egress, in the planetary frame.  The red lines represent the wavelengths of the three helium lines.}
\label{fig:excess_1D}
\end{figure}

After obtaining this residuals grid, for every column (wavelength), we subtract the mean of the out-of-transit part of the residuals image for that column; we then invert the residuals image.  The resulting residuals image now shows the excess absorption relative to the out-of-transit baseline.  However, there are still continuum variations that contribute structure in this image.  We correct for these variations by masking out the strong lines (including the helium line), fitting a 3rd order polynomial to each row (epoch) with respect to wavelength, and subtracting off the polynomial.  The result is shown in Figure \ref{fig:excess_2D}.  Finally, we shift all of the in-transit residuals spectra to the planetary rest frame and stack them to obtain the 1D excess absorption spectrum shown in Figure \ref{fig:excess_1D}.

\begin{figure}[ht]
  \includegraphics
    [width=0.5\textwidth]{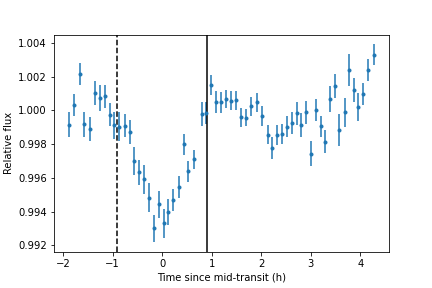}
    \caption{Light curve of the region [-20, 15] km/s from the main helium peak, in the stellar frame.  The beginning of ingress and end of egress are marked by the vertical black dotted and solid lines.  Note the stellar activity that follows the transit.}
\label{fig:light_curve}
\end{figure}

\section{Comparison to Published Observations}
\label{sec:comparison}
\begin{figure}[ht]
  \includegraphics
    [width=0.5\textwidth]{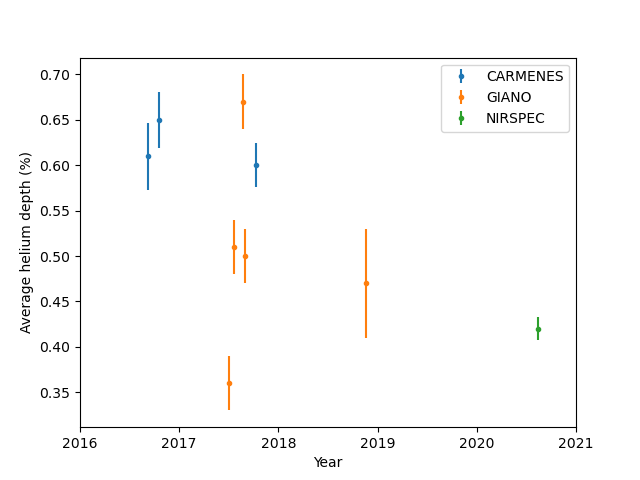}
    \caption{Comparison of the average depth of the helium line that we find (in green) to literature values, measured by CARMENES and GIANO.  All averages are computed between the end of ingress and beginning of egress in time, and between -20 and 20 km/s in planet-centric wavelength space.}
\label{fig:depth_comp}
\end{figure}

\begin{figure}[ht]
  \includegraphics
    [width=0.5\textwidth]{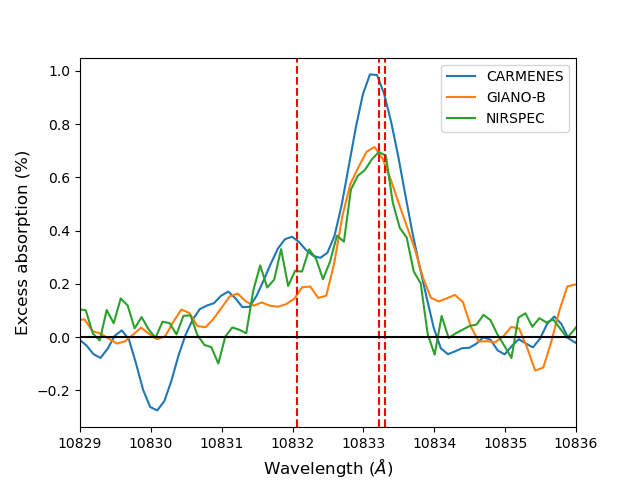}
    \caption{Excess absorption spectra obtained by the three instruments averaged over 3 transits (CARMENES), 5 transits (GIANO-B), and 1 transit (NIRSPEC).  CARMENES and GIANO-B spectra are convolved to the NIRSPEC resolution of 37,500, resulting in the artificial smoothness of the CARMENES spectrum.  These spectra are computed from the end of ingress to beginning of ingress, unlike Figure \ref{fig:excess_1D}, which is computed from the beginning of ingress to end of egress.  CARMENES and GIANO-B spectra were kindly provided by the authors of the respective papers.}
\label{fig:spec_comp}
\end{figure}

We compare our results to the eight literature values for HD 189733b's helium absorption signal. \cite{salz_2018} observed 3 transits with the fibre-fed Echelle spectrograph CARMENES (R $\sim$ 80,400), situated on the 3.5m \textit{Calar Alto} telescope.  \cite{guilluy_2020} obtained 5 transits with the slit-fed Echelle spectrograph GIANO-B (R $\sim$ 50,000) on the 3.6m \textit{Telescopio Nazionale Galileo} telescope.  The CARMENES observations have a SNR of 160--240 per spectral pixel in the continuum surrounding the helium line \citep{salz_2018}, but their exposure time is closer to five minutes than one minute, and their wavelength dispersion is 70\% higher.  Binning our observations to their time resolution and binning their observations to our spectral resolution, we find that our SNR is 1.6--2.0x times their single-night SNRs.  The same comparison, performed for the GIANO data, reveals that our SNR is 3.6--4.4x their single-night SNRs.  The GIANO comparison is not perfect, as \cite{guilluy_2020} report their SNRs over the whole order containing the helium line, not the immediate region surrounding the line.  

\cite{guilluy_2020} report the depths for each individual transit, which they calculate by averaging from the end of ingress to the beginning of egress for a wavelength range of -20 to 20 km/s in planet-centric velocity space.  We use their definition to calculate the analogous average transit depths for the CARMENES transits (using values taken from their Fig. A.3) and for our NIRSPEC transit.  We did not take into account the differing resolutions of the instruments when defining the equivalent pixel range for our wavelength bounds.  We quantified our sensitivity to the instrumental resolution by convolving the CARMENES spectrum down to the NIRSPEC resolution of 37,500 and the NIRSPEC pixel sampling before calculating the average transit depth, and found that it did not change the result by more than 0.01\% for any of the three transits.

We compare the resulting transit depths for each observation in Figure \ref{fig:depth_comp}.  This figure shows that the consistency of the three CARMENES transits over a year appears to have been a coincidence, and is not representative of the broader sample.  The GIANO transits show widely varying depths, while our NIRSPEC depth is lower than all but one of the previous transit observations.  Combining the depths for every instrument with a weighted average, we find an average transit depth of $0.617 \pm 0.017$\% for CARMENES, $0.508 \pm 0.015$\% for GIANO, and $0.420 \pm 0.013$\% for NIRSPEC.  The error bars are the standard deviation of the mean, not the sample.  The CARMENES average differs from the NIRSPEC result by more than 9$\sigma$, and from the GIANO result by 4.4$\sigma$.

\subsection{Effect of Stellar Variability on Observed He Transit Depths}
We consider two ways in which stellar variability can affect the observed transit depths.  First, if the stellar disk is inhomogenous and the planet does not transit a perfectly representative chord, the transit depth will be biased.  Second, the stellar He line may change during the observations, making the transit appear deeper or shallower than it really is.

\cite{guilluy_2020} used the five GIANO transit observations, which also included simultaneous H$\alpha$ measurements, to extensively explore the first type of variability.  H$\alpha$ is an excellent tracer of stellar activity, making it a useful proxy for the spot coverage fraction during each visit.  In active regions on the star, H$\alpha$ is seen in emission while the stellar He I line becomes deeper.  \cite{guilluy_2020} detected changes in the H$\alpha$ line during the two transit observations with the highest and lowest He I depths, suggesting that the planet was transiting an inhomogenous star.  The remaining three transits had consistent He I depths ($\sim0.5$\%) and no detectable variations in the stellar H$\alpha$ line.  The deep He I transit could be explained if the planet transited an especially quiet region of the star (thus preferentially leaving the active regions unocculted).  However, it is difficult to explain why the other transit that took place during a period of enhanced activity is shallower than expected.
\cite{guilluy_2020} argue that this observation is unlikely to be caused by a starspot occultation, as they do not observe a concurrent weakening in the Si 10830 \ang line, which is sensitive to changes in temperature.  They tentatively suggest that the planet might have occulted a filament instead.  Our new transit depth is also shallower than expected, and we also do not see any evidence for a concurrent change in the Si 10830 \ang line during the transit.  However, unlike \cite{guilluy_2020}, we do not see a mid-transit spike in the He band-integrated light curve, nor any other abnormalities that would indicate the occultation of a filament or starspot. 
 
We can quantify the effects of stellar active region contamination on the helium transmission spectrum using the techniques outlined in \citet{cauley_2018}. In order to reproduce the strength and shape of the stellar \ion{He}{1} 10833\ang triplet we found that an active region filling factor of $\approx 65\%$ and an optical depth at line center of $\tau \approx 0.7$ are required. We note that this is the facular or plage filling factor and not a spot filling factor. The large filling factor is in line with previous estimates for helium absorption on active K-dwarfs \citep{andretta_2017} and is similar to the value of $75\%$ estimated by \citet{guilluy_2020}. After estimating the filling factor, we simulated the transit of HD 189733 b across the active stellar surface and tuned the atmospheric parameters to match the observed transmission spectrum. We then tested various filling factors and geometries (e.g., uniform distribution, active latitudes) to see how the transmission spectrum changes while the atmospheric parameters are held constant. We find that even drastic changes in the filling factor ($\approx 30\%-40\%$) are not enough to reproduce the difference between the atmospheric absorption observed by \citet{salz_2018} and our NIRSPEC transmission spectrum.

The strength of the stellar helium line does not change much from epoch to epoch: \citet{guilluy_2020} found that the line core flux varied by only $\approx 4\%$ across a 1.5 year baseline. The NIRSPEC spectrum displays a similar line depth ($\approx 0.7$) compared to \citet{guilluy_2020} and \citet{salz_2018}, suggesting that the surface activity level, at least as measured by absorbing metastable helium regions, is fairly constant. This implies that the filling factor cannot change dramatically from epoch to epoch, otherwise we would have observed large variations in the stellar line strength (of order the percent change in filling factor). The relative stability of the star's filling factor and the large size of the changes in the filling factor required to account for the epoch to epoch variability in the observed helium absorption signal make it unlikely that this variability is due to inhomogeneities on the stellar surface.

On the other hand, variability in the stellar He line is a plausible contributing factor to the differing transit depths.  Figure \ref{fig:light_curve} shows the light curve of the helium line, centered slightly blueward of the main peak.  The planetary absorption is evident, but so is substantial variability after transit, including a sudden drop 2 h after mid-transit and a slow 0.6\% rise in the 2 hours following the drop.  This variability can also be seen from a close examination of Figure \ref{fig:excess_2D}, but the band-integrated light curve makes it clearer.  If similar stellar variability occurred during the transit or in the immediate out-of-transit baseline, the planetary excess absorption could appear $\sim$0.2\% larger or smaller than it actually is.

\subsection{Potential Evidence for Variability in the Planetary Outflow}

It is plausible to think that the observed variability might instead be due to changes in the planet's atmospheric outflow, which might be caused by variations in the star's high-energy radiation or stellar wind environments.  In this case, we might also expect to see visit-to-visit variations in the shape of the wavelength-dependent planetary absorption signal.  Figure \ref{fig:spec_comp} compares the excess absorption spectrum obtained by the three instruments, which tells the same story as the average depths: NIRSPEC sees a similar signal as GIANO-B (all transits combined), which in turn sees a weaker signal than CARMENES.

One interesting difference between the GIANO and NIRSPEC excess absorption spectra is that despite having a very similar primary peak, the height of the secondary peak at 10832\ang is lower for GIANO by $\sim$0.1\%.  Adopting the formal errors on the two spectra (0.07\% for GIANO and 0.05\% for NIRSPEC) and assuming statistical independence, we calculate that the difference is significant at the $\sim3.5\sigma$ level.  However, the significance of this feature is likely overestimated, as noise in high resolution spectra is often correlated across adjacent wavelengths.  If the difference is real, we can use it to constrain the optical thickness of the outflow.  A completely optically thin outflow would have a primary-to-secondary peak ratio of 8:1 (derived from the $g_i f_{ik}$ of the three lines), whereas a completely optically thick outflow would have a peak ratio of 1:1, as the gas would absorb all light in both peaks.  The higher peak ratio in the GIANO data could indicate that the absorption is coming from a more optically thick region than in our NIRSPEC observations.  However, the fact that we see significant absorption extending between the two peaks makes it difficult to determine how much of the absorption at 10832\ang is due to the line there and how much is due to absorption from strongly blueshifted gas absorbing at a rest wavelength of 10833.3\ang.  Given the right kinematic structure--namely a long tail of gas being accelerated away from the star by the stellar wind--it is conceivable that even the CARMENES observations, for which the apparent peak ratio is 2.8 \citep{salz_2018}, might be consistent with an optically thin outflow.

\section{Modeling Variability in the Planetary Outflow}
\label{sec:model}
We can use atmosphere models to quantify the effect of stellar variability on the helium absorption signal.  Here we focus on the effect of variations in the stellar XUV flux, which can be captured using relatively simple 1D hydrodynamic models.  The code we use is The PLUTO-CLOUDY Interface (TPCI) \citep{salz_2015}, a combination of two sophisticated and widely used codes: the hydrodynamic solver PLUTO \citep{mignone_2007}, and the plasma simulation and spectral synthesis code CLOUDY \citep{ferland_2013}.  TPCI has been extensively used to study photoevaporation (e.g. \citealt{salz_2015b,salz_2016,kasper_2020,zhang_2022b}).

\begin{figure}[ht]
  \includegraphics
    [width=0.5\textwidth]{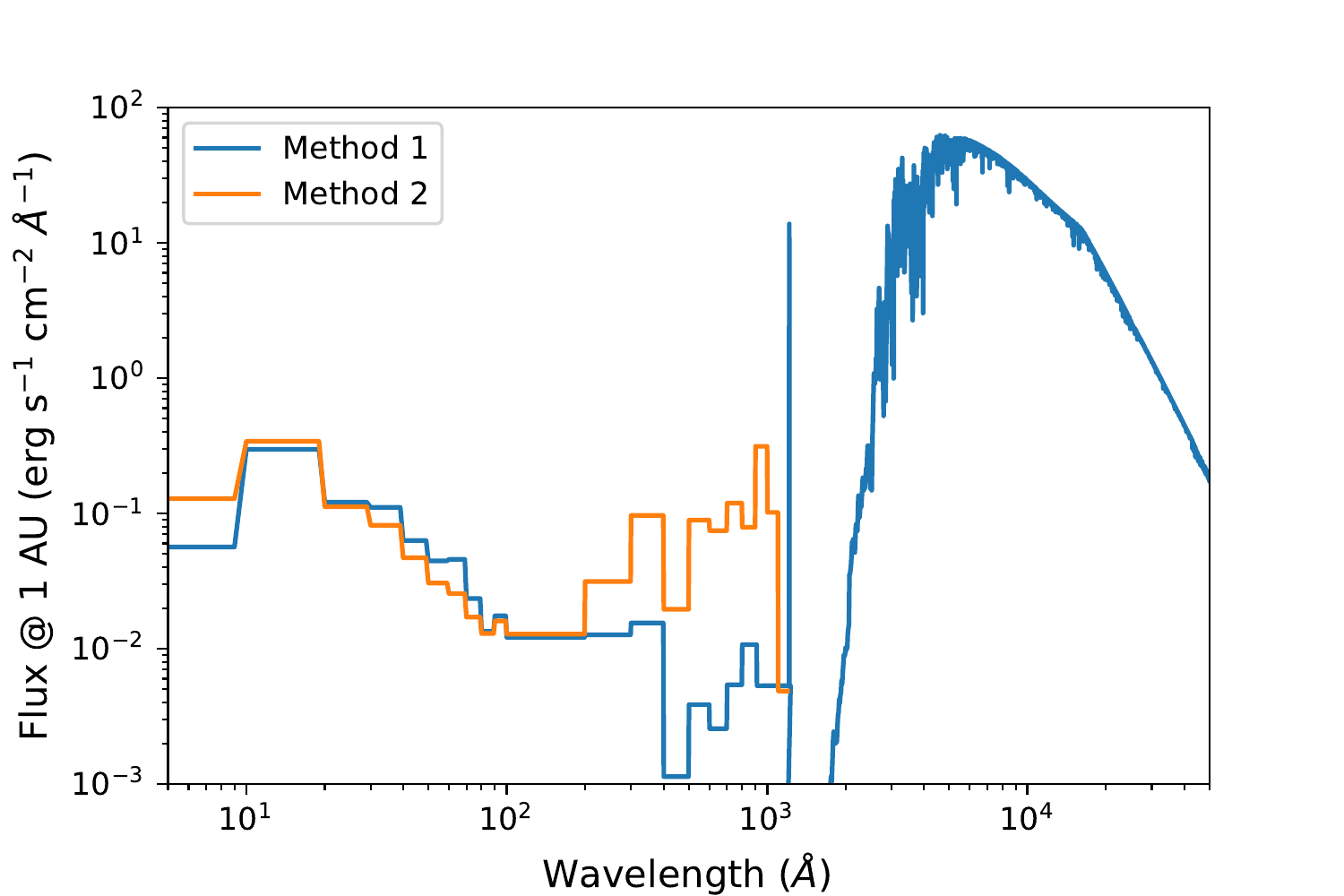}
    \caption{Reconstructed stellar spectrum using two methods.  The two methods differ blueward of 1450 \ang, where we use \cite{lampon_2021b} for method 1 and our own reconstruction for method 2.  Note the large discrepancy in the EUV.}
\label{fig:stellar_spectrum}
\end{figure}

We obtain the nominal stellar spectrum in two ways.  In the first method, we use the XUV spectrum derived by \cite{lampon_2021b} (their Fig. 2, data provided by M. Lamp{'o}n) for the wavelength range of 5--1450 \ang.  We chose a cutoff of 1450 \ang because their spectrum redward of that wavelength is not based on observations.  In the second method, the X-ray spectrum is taken from the thermal plasma model of \cite{poppenhaeger_2013}, the Ly$\alpha$ flux at 1 AU is taken to be 11.8 erg s$^{-1}$ cm$^{-2}$ \citep{france_2013}, and the EUV spectrum is derived from the Ly$\alpha$ flux using the scaling relations of \cite{linsky_2014}.  For both methods, fluxes at longer wavelengths are taken from a PHOENIX model \citep{husser_2013} ($T_{\rm eff} = 5000$, log(g)=4.5, solar metallicity).  Of particular importance is the spectrum at  1230--2588 \ang, which we call mid ultraviolet, because this radiation ionizes metastable helium, but cannot contribute to producing metastable helium because it cannot ionize ground state hydrogen.  Fortunately, we have observational constraints for the flux at these wavelengths from XMM-Newton's Optical Monitor, a photometer which operates simultaneously with the X-ray instruments.  We reviewed all archival observations taken by the OM in the UVW2 (212 nm, width 50 nm) and UVM2 (231 nm, width 48 nm) filters, and found that they do not differ by more than a few percent.  (For a summary of all XMM-Newton observations of HD 189733, see \cite{pillitteri_2022}.)  To roughly match the observations, we boosted the model spectrum between 1230--2588 \ang by 30\% and used it to predict the count rates that OM should see.  The prediction was only 6\% too low for UVW2 and 3\% too high for UVM2, indicating our model MUV flux is accurate.  The final spectra constructed using the two methods is shown in Figure \ref{fig:stellar_spectrum}, while Table \ref{table:band_fluxes} shows the band-integrated fluxes.

Comparing the two methods, the X-ray flux is in excellent agreement: \cite{lampon_2021b} reports a X-ray luminosity that corresponds to a flux of 7.8 erg s$^{-1}$ cm$^{-2}$ at 1 AU, while we obtain 7.6 erg s$^{-1}$ cm$^{-2}$.  However, the EUV (100--912 \ang) flux is dramatically discrepant.  \cite{lampon_2021b} imply a flux of 57 erg s$^{-1}$ cm$^{-2}$ at 1 AU, while we obtain 6.5 erg s$^{-1}$ cm$^{-2}$, a difference of 9x.  Other authors have obtained 11 \citep{sanz-forcada_2011}, 22 \citep{bourrier_2020}, and 36 erg s$^{-1}$ cm$^{-2}$ \citep{poppenhaeger_2013}.  This large uncertainty in the EUV flux is ultimately because the stellar EUV is only measurable from space and no space telescopes currently have EUV capabilities \citep{france_2022}.  The uncertainty has been noted by many previous publications.  For example, \cite{oklopcic_2019} compared the results of two different methods of EUV reconstruction for the same star that differed by an order of magnitude, which produced helium excess absorption depths that differed by a factor of 3.  \cite{france_2022} compared four different methods of reconstructing the EUV spectrum for Proxima Centauri and found that they were discrepant by 3--100x, depending on the wavelength.

\begin{table*}[ht]
  \centering
  \caption{Band-integrated fluxes from both stellar spectrum reconstruction methods}
  \begin{tabular}{c C C C}
  \hline
  	  Band & \text{Wavelengths} & \text{Flux (method 1)} & \text{Flux (method 2)}\\
  	  & \text{\AA} & \text{erg/s/cm}^2 & \text{erg/s/cm}^2\\
      \hline
      X-ray & 5-100 & 7.6 \pm 2.2 & 7.8 \pm 2.2\\
      EUV & 100-912 & 57 & 6.5\\
      He-ionizing & 5-512 & 24 & 12\\
      Ly$\alpha$ & 1214-1217 & 19 & 11.8 \pm 3.5 \\
      MUV & 1230-2588 & 105 \pm 7 & 105 \pm 7\\
      Total$^*$ & 5-50,000 & 4.80 \pm 0.19 \times 10^5 & 4.80 \pm 0.19 \times 10^5\\
      \hline
  \end{tabular}
  \label{table:band_fluxes}
  \\
  $^*$From \cite{addison_2019}
\end{table*}

Since the EUV flux is uncertain to an order of magnitude, we analyze the behavior of the escaping atmosphere over a wide range of EUV fluxes.  For the first stellar spectrum we constructed, which has low EUV flux compared to literature values, we set up four 1D hydrodynamic simulations: at 1x, 1.3x, 2x, and 3x the nominal stellar XUV flux as seen by the planetary dayside (which is half the flux seen by the substellar point).  For the second spectrum, which has high EUV flux compared to other literature values, we set up five 1D hydrodynamic simulations: at 0.5x, 1x, 1.3x, 2x, and 3x the nominal stellar XUV flux.  Following \cite{zhang_2022a,zhang_2022b}, we ran the simulation for 100 time units with advection off, where the time unit is calculated as the planetary radius divided by 10 km/s (roughly the sound speed).  Unlike in this previous study, we did not run the models for another 100 time units with advection on, due to numerical problems.  Also unlike in the previous study, we irradiate the atmosphere with the average flux experienced by the planetary dayside, and not the flux experienced by the substellar point.  The former is half of the latter, and is more representative of the outflow as a whole.

\begin{figure}[ht]
  \includegraphics
    [width=0.5\textwidth]{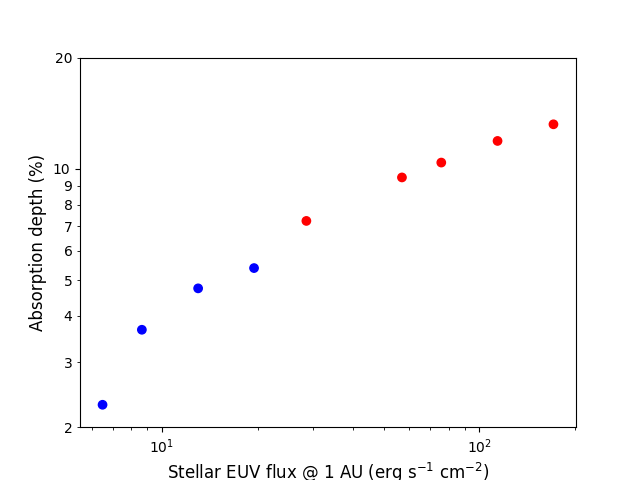}
    \caption{Relationship between maximum excess absorption depth in our 1D models and the stellar EUV flux at 1 AU.  Blue points represent XUV stellar spectra scaled from the spectrum of \cite{lampon_2021b}, while red points represent XUV spectra scaled from our own spectrum.}
\label{fig:depth_vs_XUV}
\end{figure}

Figure \ref{fig:depth_vs_XUV} shows the results of the TPCI simulations as a function of EUV flux.  The simulation results are similar to those of other 1D models.  In agreement with \cite{lampon_2021a}, we find that the the outflow is hot ($>\sim$10,000 K) and in the recombination regime, which is characterized by a sharp transition from neutral to ionized hydrogen.  For the 1/2 nominal XUV model using the first stellar spectrum, which is the model with median EUV flux (see Figure \ref{fig:depth_vs_XUV}), the outflow is 95\% neutral at 1.12 $R_p$ and drops to 5\% neutral by 1.32 $R_p$.  In the recombination regime, most of the incident XUV is radiated away via recombination, leading to low mass loss efficiency.  In fact, the mass loss rate we obtain for the aforementioned model, $5 \times 10^9$ g/s, is substantially lower than even the $10^{11}$ g/s found by \cite{lampon_2021a,lampon_2021b}, and similar to the $10^{10}$ g/s found by \cite{caldiroli_2022} for similar XUV flux.  This corresponds to a mass loss efficiency of 1.3\%, similar to the 2--3\% found by \cite{caldiroli_2022}.

The absorption depth is higher than observed even in the simulation with the lowest EUV.  Increasing the EUV flux 34\% above this lowest point increased the helium absorption by 60\%, but the depth increases roughly with the square root of EUV flux thereafter, so that further 34\% increases in flux only produce 15\% increases in depth.  

HD 189733 has been extensively observed at XUV wavelengths, providing us with good empirical constraints on the magnitude of the star's variability. \cite{pillitteri_2022} analyzed 25 XMM-Newton observations spanning a total of 8 years, the last of which was taken before the first helium observation.  They found that the 25--75\% range of quiescent flux varies by 23\%, and that 44\% of the total observing time was occupied by flares, which typically lasted several kiloseconds and increased the XUV flux by a few tens of percent (see their Fig. B.2).  Using the data behind their Figure B.3 (Pillitteri, private comm.), we calculate that the typical variability in the X-ray flux, as defined by the gap between the 84th and 16th percentiles, is 60\%.  EUV flux increases with X-ray flux as $F_{\rm EUV} \propto F_{\rm X}^{0.86}$ \citep{sanz-forcada_2011}, so a 60\% increase in X-rays corresponds to a 50\% increase in EUV, and a 23\% increase in X-rays corresponds to a 19\% increase in EUV. These numbers suggest that the observational variability plotted in Figure \ref{fig:depth_comp} can plausibly be due to variability in stellar EUV output if the EUV flux is low ($\sim$6 erg s$^{-1}$ cm$^{-2}$ at 1 AU), but is less likely if the EUV flux is high ($>\sim$10).

It is worthwhile to briefly consider other sources of variability that we could not explore with our 1D model.  The effect of differing stellar wind conditions on the outflow has been extensively studied with 3D simulations (e.g. \citealt{mccann_2019,vidotto_2020}), but generally without modelling the helium line.  \cite{macleod_2022} does model the helium line, and find that given the same planetary mass loss rate, the equivalent width of helium absorption changes by 10\% when the stellar wind is increased 10x from weak to moderate, but changes far more drastically when the wind is strong enough to ``break through'' the sonic surface and confine the outflow.  \cite{zhang_2022b} study the effect of different stellar wind conditions on the helium absorption from TOI 560b, a young mini-Neptune, by alternately halving the stellar wind density and velocity.  They find changes in peak absorption of up to 25\%, but smaller changes in equivalent width of up to 12\%.  These studies show that changes in stellar wind conditions can plausibly explain a portion of the variability we see in HD 189733b.

Another source of variability is shear instability, which exists even with constant irradiation and wind from the star.  For the inflated sub-Saturn WASP-107b, \cite{wang_2021a} uses a 3D hydrodynamic model to predict $\sim$10\% fluctuations in helium absorption over hour-long timescales.  These fluctuations, however, change the planetary mass loss rate by less than 1\%.  Shear instability, like the impact of the stellar wind, is not possible to model in 1D.

One final source of variability that has been modelled in the literature is flares.  Flares increase the star's XUV output, and a sufficiently long-lasting flare would increase helium absorption, as predicted by our models.  However, depending on the spectrum of the flare, the immediate effect of a flare might be to ionize metastable helium and reduce helium absorption.  After a dynamical timescale, when the surge of photoevaporative mass loss generated by the flare has reached higher altitudes, helium absorption should increases above baseline \citep{wang_2021b}.  For WASP-69b, the planet simulated by \cite{wang_2021b}, this takes about an hour; helium absorption then stays elevated for 2 days after the end of the flare.  Thus, even though flares can both decrease and increase helium absorption, the latter is far more likely to be observed than the former.  To explore the effect of flares that raise the XUV flux by tens of percent on a high-gravity planet like HD 189733b--rather than a 10x flare on a low gravity planet like WASP-69b, as \cite{wang_2021b} did--new 3D simulations are desirable.


\section{Conclusion}
\label{sec:conclusion}
Metastable helium observations provide us with a sensitive probe of planetary outflow characteristics.  In this paper, we present a new helium transit observation of HD 189733b, a hot Jupiter that has been observed over a wide range of wavelengths in the fifteen years since its discovery.  This planet will be also observed by 5 \emph{James Webb Space Telescope} programs in Cycle 1 alone; these programs will doubtlessly further expand our knowledge of its atmospheric properties.

When combined with the set of eight published helium transit observations of HD 189733b, our new observations add to the growing body of evidence suggesting that this planet's outflow properties may vary in time.  HD 189733 is a relatively active K dwarf and it has been extensively observed at high energies by XMM-Newton and Chandra.  Our simulations suggest that stellar XUV variability, by itself, can plausibly explain the observed variations in the planet's helium transit depth.  However, this does not mean that other sources of variability we do not model, such as stellar wind variations, shear instability, and short flares, cannot also explain the variability.

Understanding the high-energy environment and hydrostatic atmosphere of the planet is crucial to understanding mass loss, and the quantity of atmospheric observations available for HD 189733b make it one of the most promising planets for understanding mass loss physics in detail.  In order to better understand the amplitude, timescale, and likely cause of this planet's observed variability, it is important to continue monitoring of the planet in the helium line, preferably with simultaneous H$\alpha$ and/or X-ray measurements.  This will in turn inform how much we should rely on single-epoch observations of helium absorption to quantify the mass loss rates and population-level properties of other close-in planets.

\textit{Software:}  \texttt{numpy \citep{van_der_walt_2011}, scipy \citep{virtanen_2020}, matplotlib \citep{hunter_2007}}

\acknowledgments
The data presented herein were obtained at the W. M. Keck Observatory, which is operated as a scientific partnership among the California Institute of Technology, the University of California and the National Aeronautics and Space Administration. The Observatory was made possible by the generous financial support of the W. M. Keck Foundation.  AO gratefully acknowledges support from the Dutch Research Council NWO Veni grant.

\bibliographystyle{apj} \bibliography{main}

\begin{thebibliography}{54}
\expandafter\ifx\csname natexlab\endcsname\relax\def\natexlab#1{#1}\fi

\bibitem[{{Adams}(2011)}]{adams_2011}
{Adams}, F.~C. 2011, \apj, 730, 27

\bibitem[{{Addison} {et~al.}(2019){Addison}, {Wright}, {Wittenmyer}, {Horner},
  {Mengel}, {Johns}, {Marti}, {Nicholson}, {Soutter}, {Bowler}, {Crossfield},
  {Kane}, {Kielkopf}, {Plavchan}, {Tinney}, {Zhang}, {Clark}, {Clerte},
  {Eastman}, {Swift}, {Bottom}, {Muirhead}, {McCrady}, {Herzig}, {Hogstrom},
  {Wilson}, {Sliski}, {Johnson}, {Wright}, {Johnson}, {Blake}, {Riddle}, {Lin},
  {Cornachione}, {Bedding}, {Stello}, {Huber}, {Marsden}, \&
  {Carter}}]{addison_2019}
{Addison}, B., {Wright}, D.~J., {Wittenmyer}, R.~A., {et~al.} 2019, \pasp, 131,
  115003

\bibitem[{{Allart} {et~al.}(2019){Allart}, {Bourrier}, {Lovis}, {Ehrenreich},
  {Aceituno}, {Guijarro}, {Pepe}, {Sing}, {Spake}, \&
  {Wyttenbach}}]{allart_2019}
{Allart}, R., {Bourrier}, V., {Lovis}, C., {et~al.} 2019, \aap, 623, A58

\bibitem[{{Andretta} {et~al.}(2017){Andretta}, {Giampapa}, {Covino}, {Reiners},
  \& {Beeck}}]{andretta_2017}
{Andretta}, V., {Giampapa}, M.~S., {Covino}, E., {Reiners}, A., \& {Beeck}, B.
  2017, \apj, 839, 97

\bibitem[{{Bouchy} {et~al.}(2005){Bouchy}, {Udry}, {Mayor}, {Moutou}, {Pont},
  {Iribarne}, {da Silva}, {Ilovaisky}, {Queloz}, {Santos}, {S{\'e}gransan}, \&
  {Zucker}}]{bouchy_2005}
{Bouchy}, F., {Udry}, S., {Mayor}, M., {et~al.} 2005, \aap, 444, L15

\bibitem[{Bourrier {et~al.}(2020)Bourrier, Wheatley, Lecavelier des Etangs,
  King, Louden, Ehrenreich, Fares, Helling, Llama, Jardine, \&
  Vidotto}]{bourrier_2020}
Bourrier, V., Wheatley, P.~J., Lecavelier des Etangs, A., {et~al.} 2020,
  Monthly Notices of the Royal Astronomical Society, 493, 559

\bibitem[{{Caldiroli} {et~al.}(2022){Caldiroli}, {Haardt}, {Gallo}, {Spinelli},
  {Malsky}, \& {Rauscher}}]{caldiroli_2022}
{Caldiroli}, A., {Haardt}, F., {Gallo}, E., {et~al.} 2022, \aap, 663, A122

\bibitem[{{Cauley} {et~al.}(2018){Cauley}, {Kuckein}, {Redfield}, {Shkolnik},
  {Denker}, {Llama}, \& {Verma}}]{cauley_2018}
{Cauley}, P.~W., {Kuckein}, C., {Redfield}, S., {et~al.} 2018, \aj, 156, 189

\bibitem[{{Ferland} {et~al.}(2013){Ferland}, {Porter}, {van Hoof}, {Williams},
  {Abel}, {Lykins}, {Shaw}, {Henney}, \& {Stancil}}]{ferland_2013}
{Ferland}, G.~J., {Porter}, R.~L., {van Hoof}, P.~A.~M., {et~al.} 2013, \rmxaa,
  49, 137

\bibitem[{{France} {et~al.}(2013){France}, {Froning}, {Linsky}, {Roberge},
  {Stocke}, {Tian}, {Bushinsky}, {D{\'e}sert}, {Mauas}, {Vieytes}, \&
  {Walkowicz}}]{france_2013}
{France}, K., {Froning}, C.~S., {Linsky}, J.~L., {et~al.} 2013, \apj, 763, 149

\bibitem[{{France} {et~al.}(2022){France}, {Fleming}, {Youngblood}, {Mason},
  {Drake}, {Amerstorfer}, {Barstow}, {Bourrier}, {Champey}, {Fossati},
  {Froning}, {Green}, {Gris{\'e}}, {Gronoff}, {Hellickson}, {Jin}, {Koskinen},
  {Kowalski}, {Kruczek}, {Linsky}, {Lipscy}, {McEntaffer}, {McKenzie}, {Miles},
  {Patton}, {Savage}, {Siegmund}, {Spittler}, {Unruh}, \& {Volz}}]{france_2022}
{France}, K., {Fleming}, B., {Youngblood}, A., {et~al.} 2022, Journal of
  Astronomical Telescopes, Instruments, and Systems, 8, 014006

\bibitem[{{Fulton} \& {Petigura}(2018)}]{fulton_2018}
{Fulton}, B.~J., \& {Petigura}, E.~A. 2018, \aj, 156, 264

\bibitem[{{Fulton} {et~al.}(2017){Fulton}, {Petigura}, {Howard}, {Isaacson},
  {Marcy}, {Cargile}, {Hebb}, {Weiss}, {Johnson}, {Morton}, {Sinukoff},
  {Crossfield}, \& {Hirsch}}]{fulton_2017}
{Fulton}, B.~J., {Petigura}, E.~A., {Howard}, A.~W., {et~al.} 2017, \aj, 154,
  109

\bibitem[{{Guilluy} {et~al.}(2020){Guilluy}, {Andretta}, {Borsa}, {Giacobbe},
  {Sozzetti}, {Covino}, {Bourrier}, {Fossati}, {Bonomo}, {Esposito},
  {Giampapa}, {Harutyunyan}, {Rainer}, {Brogi}, {Bruno}, {Claudi}, {Frustagli},
  {Lanza}, {Mancini}, {Pino}, {Poretti}, {Scandariato}, {Affer}, {Baffa},
  {Baruffolo}, {Benatti}, {Biazzo}, {Bignamini}, {Boschin}, {Carleo},
  {Cecconi}, {Cosentino}, {Damasso}, {Desidera}, {Falcini}, {Martinez
  Fiorenzano}, {Ghedina}, {Gonz{\'a}lez-{\'A}lvarez}, {Guerra}, {Hernandez},
  {Leto}, {Maggio}, {Malavolta}, {Maldonado}, {Micela}, {Molinari},
  {Nascimbeni}, {Pagano}, {Pedani}, {Piotto}, \& {Reiners}}]{guilluy_2020}
{Guilluy}, G., {Andretta}, V., {Borsa}, F., {et~al.} 2020, \aap, 639, A49

\bibitem[{{Harrington} {et~al.}(2022){Harrington}, {Himes}, {Cubillos},
  {Blecic}, {Rojo}, {Challener}, {Lust}, {Bowman}, {Blumenthal}, {Dobbs-Dixon},
  {Foster}, {Foster}, {Green}, {Loredo}, {McIntyre}, {Stemm}, \&
  {Wright}}]{harrington_2022}
{Harrington}, J., {Himes}, M.~D., {Cubillos}, P.~E., {et~al.} 2022, The
  Planetary Science Journal, 3, 80

\bibitem[{{Hunter}(2007)}]{hunter_2007}
{Hunter}, J.~D. 2007, Computing in Science and Engineering, 9, 90

\bibitem[{{Husser} {et~al.}(2013){Husser}, {Wende-von Berg}, {Dreizler},
  {Homeier}, {Reiners}, {Barman}, \& {Hauschildt}}]{husser_2013}
{Husser}, T.-O., {Wende-von Berg}, S., {Dreizler}, S., {et~al.} 2013, \aap,
  553, A6

\bibitem[{{Kasper} {et~al.}(2020){Kasper}, {Bean}, {Oklop{\v{c}}i{\'c}},
  {Malsky}, {Kempton}, {D{\'e}sert}, {Rogers}, \& {Mansfield}}]{kasper_2020}
{Kasper}, D., {Bean}, J.~L., {Oklop{\v{c}}i{\'c}}, A., {et~al.} 2020, \aj, 160,
  258

\bibitem[{{Knutson} {et~al.}(2007){Knutson}, {Charbonneau}, {Allen}, {Fortney},
  {Agol}, {Cowan}, {Showman}, {Cooper}, \& {Megeath}}]{knutson_2007}
{Knutson}, H.~A., {Charbonneau}, D., {Allen}, L.~E., {et~al.} 2007, \nat, 447,
  183

\bibitem[{{Lamp{\'o}n} {et~al.}(2021{\natexlab{a}}){Lamp{\'o}n},
  {L{\'o}pez-Puertas}, {Czesla}, {S{\'a}nchez-L{\'o}pez}, {Lara}, {Salz},
  {Sanz-Forcada}, {Molaverdikhani}, {Quirrenbach}, {Pall{\'e}}, {Caballero},
  {Henning}, {Nortmann}, {Amado}, {Montes}, {Reiners}, \&
  {Ribas}}]{lampon_2021a}
{Lamp{\'o}n}, M., {L{\'o}pez-Puertas}, M., {Czesla}, S., {et~al.}
  2021{\natexlab{a}}, \aap, 648, L7

\bibitem[{{Lamp{\'o}n} {et~al.}(2021{\natexlab{b}}){Lamp{\'o}n},
  {L{\'o}pez-Puertas}, {Sanz-Forcada}, {S{\'a}nchez-L{\'o}pez},
  {Molaverdikhani}, {Czesla}, {Quirrenbach}, {Pall{\'e}}, {Caballero},
  {Henning}, {Salz}, {Nortmann}, {Aceituno}, {Amado}, {Bauer}, {Montes},
  {Nagel}, {Reiners}, \& {Ribas}}]{lampon_2021b}
{Lamp{\'o}n}, M., {L{\'o}pez-Puertas}, M., {Sanz-Forcada}, J., {et~al.}
  2021{\natexlab{b}}, \aap, 647, A129

\bibitem[{{Linsky} {et~al.}(2014){Linsky}, {Fontenla}, \&
  {France}}]{linsky_2014}
{Linsky}, J.~L., {Fontenla}, J., \& {France}, K. 2014, \apj, 780, 61

\bibitem[{{MacLeod} \& {Oklop{\v{c}}i{\'c}}(2022)}]{macleod_2022}
{MacLeod}, M., \& {Oklop{\v{c}}i{\'c}}, A. 2022, \apj, 926, 226

\bibitem[{McCann {et~al.}(2019)McCann, Murray-Clay, Kratter, \&
  Krumholz}]{mccann_2019}
McCann, J., Murray-Clay, R.~A., Kratter, K., \& Krumholz, M.~R. 2019, The
  Astrophysical Journal, 873, 89

\bibitem[{{McCullough} {et~al.}(2014){McCullough}, {Crouzet}, {Deming}, \&
  {Madhusudhan}}]{mccullough_2014}
{McCullough}, P.~R., {Crouzet}, N., {Deming}, D., \& {Madhusudhan}, N. 2014,
  \apj, 791, 55

\bibitem[{Mignone {et~al.}(2007)Mignone, Bodo, Massaglia, Matsakos, Tesileanu,
  Zanni, \& Ferrari}]{mignone_2007}
Mignone, A., Bodo, G., Massaglia, S., {et~al.} 2007, The Astrophysical Journal
  Supplement Series, 170, 228

\bibitem[{{Morello} {et~al.}(2014){Morello}, {Waldmann}, {Tinetti}, {Peres},
  {Micela}, \& {Howarth}}]{morello_2014}
{Morello}, G., {Waldmann}, I.~P., {Tinetti}, G., {et~al.} 2014, \apj, 786, 22

\bibitem[{{Nortmann} {et~al.}(2018){Nortmann}, {Pall{\'e}}, {Salz},
  {Sanz-Forcada}, {Nagel}, {Alonso-Floriano}, {Czesla}, {Yan}, {Chen},
  {Snellen}, {Zechmeister}, {Schmitt}, {L{\'o}pez-Puertas}, {Casasayas-Barris},
  {Bauer}, {Amado}, {Caballero}, {Dreizler}, {Henning}, {Lamp{\'o}n}, {Montes},
  {Molaverdikhani}, {Quirrenbach}, {Reiners}, {Ribas}, {S{\'a}nchez-L{\'o}pez},
  {Schneider}, \& {Zapatero Osorio}}]{nortmann_2018}
{Nortmann}, L., {Pall{\'e}}, E., {Salz}, M., {et~al.} 2018, Science, 362, 1388

\bibitem[{{Oklop{\v{c}}i{\'c}}(2019)}]{oklopcic_2019}
{Oklop{\v{c}}i{\'c}}, A. 2019, \apj, 881, 133

\bibitem[{{Oklop{\v{c}}i{\'c}} \& {Hirata}(2018)}]{oklopcic_2018}
{Oklop{\v{c}}i{\'c}}, A., \& {Hirata}, C.~M. 2018, \apj, 855, L11

\bibitem[{{Pillitteri} {et~al.}(2022){Pillitteri}, {Micela}, {Maggio},
  {Sciortino}, \& {Lopez-Santiago}}]{pillitteri_2022}
{Pillitteri}, I., {Micela}, G., {Maggio}, A., {Sciortino}, S., \&
  {Lopez-Santiago}, J. 2022, \aap, 660, A75

\bibitem[{{Poppenhaeger}(2022)}]{poppenhaeger_2022}
{Poppenhaeger}, K. 2022, \mnras, 512, 1751

\bibitem[{{Poppenhaeger} {et~al.}(2013){Poppenhaeger}, {Schmitt}, \&
  {Wolk}}]{poppenhaeger_2013}
{Poppenhaeger}, K., {Schmitt}, J.~H.~M.~M., \& {Wolk}, S.~J. 2013, \apj, 773,
  62

\bibitem[{{Rosenthal} {et~al.}(2021){Rosenthal}, {Fulton}, {Hirsch},
  {Isaacson}, {Howard}, {Dedrick}, {Sherstyuk}, {Blunt}, {Petigura}, {Knutson},
  {Behmard}, {Chontos}, {Crepp}, {Crossfield}, {Dalba}, {Fischer}, {Henry},
  {Kane}, {Kosiarek}, {Marcy}, {Rubenzahl}, {Weiss}, \&
  {Wright}}]{rosenthal_2021}
{Rosenthal}, L.~J., {Fulton}, B.~J., {Hirsch}, L.~A., {et~al.} 2021, \apjs,
  255, 8

\bibitem[{{Salz} {et~al.}(2015{\natexlab{a}}){Salz}, {Banerjee}, {Mignone},
  {Schneider}, {Czesla}, \& {Schmitt}}]{salz_2015}
{Salz}, M., {Banerjee}, R., {Mignone}, A., {et~al.} 2015{\natexlab{a}}, \aap,
  576, A21

\bibitem[{{Salz} {et~al.}(2016){Salz}, {Czesla}, {Schneider}, \&
  {Schmitt}}]{salz_2016}
{Salz}, M., {Czesla}, S., {Schneider}, P.~C., \& {Schmitt}, J.~H.~M.~M. 2016,
  \aap, 586, A75

\bibitem[{{Salz} {et~al.}(2015{\natexlab{b}}){Salz}, {Schneider}, {Czesla}, \&
  {Schmitt}}]{salz_2015b}
{Salz}, M., {Schneider}, P.~C., {Czesla}, S., \& {Schmitt}, J.~H.~M.~M.
  2015{\natexlab{b}}, \aap, 576, A42

\bibitem[{{Salz} {et~al.}(2018){Salz}, {Czesla}, {Schneider}, {Nagel},
  {Schmitt}, {Nortmann}, {Alonso-Floriano}, {L{\'o}pez-Puertas}, {Lamp{\'o}n},
  {Bauer}, {Snellen}, {Pall{\'e}}, {Caballero}, {Yan}, {Chen}, {Sanz-Forcada},
  {Amado}, {Quirrenbach}, {Ribas}, {Reiners}, {B{\'e}jar}, {Casasayas-Barris},
  {Cort{\'e}s-Contreras}, {Dreizler}, {Guenther}, {Henning}, {Jeffers},
  {Kaminski}, {K{\"u}rster}, {Lafarga}, {Lara}, {Molaverdikhani}, {Montes},
  {Morales}, {S{\'a}nchez-L{\'o}pez}, {Seifert}, {Zapatero Osorio}, \&
  {Zechmeister}}]{salz_2018}
{Salz}, M., {Czesla}, S., {Schneider}, P.~C., {et~al.} 2018, \aap, 620, A97

\bibitem[{{Sanz-Forcada} {et~al.}(2011){Sanz-Forcada}, {Micela}, {Ribas},
  {Pollock}, {Eiroa}, {Velasco}, {Solano}, \&
  {Garc{\'\i}a-{\'A}lvarez}}]{sanz-forcada_2011}
{Sanz-Forcada}, J., {Micela}, G., {Ribas}, I., {et~al.} 2011, \aap, 532, A6

\bibitem[{{Seager} \& {Sasselov}(2000)}]{seager_2000}
{Seager}, S., \& {Sasselov}, D.~D. 2000, \apj, 537, 916

\bibitem[{Showman {et~al.}(2009)Showman, Fortney, Lian, Marley, Freedman,
  Knutson, \& Charbonneau}]{showman_2009}
Showman, A.~P., Fortney, J.~J., Lian, Y., {et~al.} 2009, The Astrophysical
  Journal, 699, 564

\bibitem[{{Sing} {et~al.}(2011){Sing}, {Pont}, {Aigrain}, {Charbonneau},
  {D{\'e}sert}, {Gibson}, {Gilliland}, {Hayek}, {Henry}, {Knutson}, {Lecavelier
  Des Etangs}, {Mazeh}, \& {Shporer}}]{sing_2011}
{Sing}, D.~K., {Pont}, F., {Aigrain}, S., {et~al.} 2011, \mnras, 416, 1443

\bibitem[{{Spake} {et~al.}(2018){Spake}, {Sing}, {Evans}, {Oklop{\v{c}}i{\'c}},
  {Bourrier}, {Kreidberg}, {Rackham}, {Irwin}, {Ehrenreich}, {Wyttenbach},
  {Wakeford}, {Zhou}, {Chubb}, {Nikolov}, {Goyal}, {Henry}, {Williamson},
  {Blumenthal}, {Anderson}, {Hellier}, {Charbonneau}, {Udry}, \&
  {Madhusudhan}}]{spake_2018}
{Spake}, J.~J., {Sing}, D.~K., {Evans}, T.~M., {et~al.} 2018, \nat, 557, 68

\bibitem[{{Steinrueck} {et~al.}(2021){Steinrueck}, {Showman}, {Lavvas},
  {Koskinen}, {Tan}, \& {Zhang}}]{steinrueck_2021}
{Steinrueck}, M.~E., {Showman}, A.~P., {Lavvas}, P., {et~al.} 2021, \mnras,
  504, 2783

\bibitem[{Szab{\'{o}} \& Kiss(2011)}]{szab_2011}
Szab{\'{o}}, G.~M., \& Kiss, L.~L. 2011, The Astrophysical Journal, 727, L44

\bibitem[{{van der Walt} {et~al.}(2011){van der Walt}, {Colbert}, \&
  {Varoquaux}}]{van_der_walt_2011}
{van der Walt}, S., {Colbert}, S.~C., \& {Varoquaux}, G. 2011, Computing in
  Science and Engineering, 13, 22

\bibitem[{Vidotto \& Cleary(2020)}]{vidotto_2020}
Vidotto, A.~A., \& Cleary, A. 2020, Monthly Notices of the Royal Astronomical
  Society, 494, 2417

\bibitem[{{Virtanen} {et~al.}(2020){Virtanen}, {Gommers}, {Oliphant},
  {Haberland}, {Reddy}, {Cournapeau}, {Burovski}, {Peterson}, {Weckesser},
  {Bright}, {van der Walt}, {Brett}, {Wilson}, {Jarrod Millman}, {Mayorov},
  {Nelson}, {Jones}, {Kern}, {Larson}, {Carey}, {Polat}, {Feng}, {Moore}, {Vand
  erPlas}, {Laxalde}, {Perktold}, {Cimrman}, {Henriksen}, {Quintero}, {Harris},
  {Archibald}, {Ribeiro}, {Pedregosa}, {van Mulbregt}, \&
  {Contributors}}]{virtanen_2020}
{Virtanen}, P., {Gommers}, R., {Oliphant}, T.~E., {et~al.} 2020, Nature
  Methods, 17, 261

\bibitem[{{Wang} \& {Dai}(2021{\natexlab{a}})}]{wang_2021b}
{Wang}, L., \& {Dai}, F. 2021{\natexlab{a}}, \apj, 914, 98

\bibitem[{{Wang} \& {Dai}(2021{\natexlab{b}})}]{wang_2021a}
---. 2021{\natexlab{b}}, \apj, 914, 99

\bibitem[{{Zhang} {et~al.}(2020){Zhang}, {Chachan}, {Kempton}, {Knutson}, \&
  {Chang}}]{zhang_2020b}
{Zhang}, M., {Chachan}, Y., {Kempton}, E. M.~R., {Knutson}, H.~A., \& {Chang},
  W.~H. 2020, \apj, 899, 27

\bibitem[{{Zhang} {et~al.}(2022{\natexlab{a}}){Zhang}, {Knutson}, {Wang},
  {Dai}, \& {Barrag{\'a}n}}]{zhang_2022b}
{Zhang}, M., {Knutson}, H.~A., {Wang}, L., {Dai}, F., \& {Barrag{\'a}n}, O.
  2022{\natexlab{a}}, \aj, 163, 67

\bibitem[{{Zhang} {et~al.}(2021){Zhang}, {Knutson}, {Wang}, {Dai}, {Oklopcic},
  \& {Hu}}]{zhang_2021}
{Zhang}, M., {Knutson}, H.~A., {Wang}, L., {et~al.} 2021, \aj, 161, 181

\bibitem[{{Zhang} {et~al.}(2022{\natexlab{b}}){Zhang}, {Knutson}, {Wang},
  {Dai}, {dos Santos}, {Fossati}, {Henry}, {Ehrenreich}, {Alibert}, {Hoyer},
  {Wilson}, \& {Bonfanti}}]{zhang_2022a}
---. 2022{\natexlab{b}}, \aj, 163, 68

\end{thebibliography}

\end{document}